\documentclass[pra,10pt,showpacs,amsmath,twocolumn,floatfix]{revtex4}
\usepackage{amsmath}
\usepackage{amsfonts}
\usepackage{graphicx}

\newcommand{\parti}[2]{\frac{\partial #1}{\partial #2}}
\newcommand{\partit}[2]{\frac{\partial^2 #1}{\partial #2^2}}
\newcommand{\diff}[2]{\frac{d #1}{d #2}}

\newcommand{\intall}{\int_{-\infty}^{\infty}}
\newcommand{\ket}[1]{|#1\rangle}

\newcommand{\bra}[1]{\langle#1|}

\newcommand{\bs}[1]{\boldsymbol{#1}}

\newcommand{\bk}[1]{\left(#1\right)}
\newcommand{\Bk}[1]{\left[#1\right]}
\newcommand{\BK}[1]{\left\{#1\right\}}

\newcommand{\trace}{\operatorname{tr}}

\begin{document}
\title{Continuous Quantum Hypothesis Testing}

\author{Mankei Tsang}

\email{eletmk@nus.edu.sg}
\affiliation{Department of Electrical and Computer Engineering,
  National University of Singapore, 4 Engineering Drive 3, Singapore
  117583}

\affiliation{Department of Physics, National University of Singapore,
  2 Science Drive 3, Singapore 117551}









\date{\today}

\begin{abstract}
  I propose a general quantum hypothesis testing theory that enables
  one to test hypotheses about any aspect of a physical system,
  including its dynamics, based on a series of observations. For
  example, the hypotheses can be about the presence of a weak
  classical signal continuously coupled to a quantum sensor, or about
  competing quantum or classical models of the dynamics of a system.
  This generalization makes the theory useful for quantum detection
  and experimental tests of quantum mechanics in general.  In the case
  of continuous measurements, the theory is significantly simplified
  to produce compact formulae for the likelihood ratio, the central
  quantity in statistical hypothesis testing.  The likelihood ratio
  can then be computed efficiently in many cases of interest. Two
  potential applications of the theory, namely quantum detection of a
  classical stochastic waveform and test of harmonic-oscillator energy
  quantization, are discussed.
 \end{abstract}

\maketitle
Testing hypotheses about a physical system by observation is a
fundamental endeavor in scientific research. Observations are often
indirect, noisy, and limited; to choose the best model of a system
among potential candidates, statistical inference is the most logical
way \cite{jaynes,vantrees} and has been extensively employed in
diverse fields of science and engineering.

Many important quantum mechanics experiments, such as tests of quantum
mechanics \cite{bell,bednorz}, quantum detection of weak forces or
magnetic fields \cite{braginsky}, and quantum target detection
\cite{helstrom,target}, are examples of hypothesis testing. To test
quantum nonlocality, for instance, one should compare the quantum
model with the best classical model; Bell's inequality and its
variations, which impose general bounds on observations of
local-hidden-variable systems, have been widely used in this regard
\cite{bell,bednorz}. The analyses of experimental data in many such
tests have nonetheless been criticized by Peres \cite{peres}: The
statistical averages in all these inequalities can never be measured
exactly in a finite number of trials. One should use statistical
inference to account for the uncertainties and provide an operational
meaning to the data.

Another important recent development in quantum physics is the
experimental demonstration of quantum behavior in increasingly
macroscopic systems, such as mechanical oscillators
\cite{mechanical,thompson} and microwave resonators
\cite{electrical}. To test the quantization of the oscillator energy
\cite{thompson,electrical}, for example, the use of quantum filtering
theory has been proposed to process the data \cite{santamore}, but
testing quantum behavior by assuming quantum mechanics can be
criticized as begging the question. An ingenious proposal by Clerk
\textit{et al.}\ considers the third moment of energy as a test of
energy quantization \cite{clerk}. Like the correlations in Bell's
inequality, however, the third moment is a statistical average and
cannot be measured exactly in finite time. Again, statistical
inference should be used to test the quantum behavior of a system
rigorously, especially when the measurements are weak and noisy. The
good news here is that the error probabilities for hypothesis testing
should decrease exponentially with the number of measurements when the
number is large \cite{vantrees}, so one can always compensate for a
weak signal-to-noise ratio by increasing the number of trials.

Quantum hypothesis testing was first studied by Holevo \cite{holevo}
and Yuen \textit{et al.}\ \cite{yuen}. Since then, researchers have
focused on the use of statistical hypothesis testing techniques for
initial quantum state discrimination \cite{helstrom,chefles}.  Here I
propose a more general quantum theory of hypothesis testing for
\emph{model} discrimination, allowing the hypotheses to be not just
about the initial state but also about the dynamics of the system
under a series of observations.  This generalization makes the theory
applicable to virtually any hypothesis testing problem that involves
quantum mechanics, including tests of quantum dynamics \cite{bednorz}
and quantum waveform detection \cite{braginsky}.

In the case of continuous measurements with Gaussian or Poissonian
noise, the theory is significantly simplified to produce compact
formulae for the likelihood ratio, the central quantity in statistical
hypothesis testing. The formulae enable one to compute the ratio
efficiently in many cases of interest and should be useful for
numerical approximations in general. Notable prior work on continuous
quantum hypothesis testing is reported in Refs.~\cite{gambetta,chase},
which study state discrimination or parameter estimation only and have
not derived the general likelihood-ratio formulae proposed here.

To illustrate the theory, I discuss two potential applications, namely
quantum detection of a classical stochastic waveform and test of
harmonic-oscillator energy quantization.  Waveform detection is a
basic operation in future quantum sensing applications, such as
gravitational-wave detection, optomechanical force detection, and
atomic magnetometry \cite{braginsky}. Tests of energy quantization, on
the other hand, have become increasingly popular in experimental
physics due to the rapid recent progress in device fabrication
technologies \cite{mechanical,thompson,electrical}. Besides these two
applications, the theory is expected to find wide use in quantum
information processing and quantum physics in general, whenever new
claims about a quantum system need to be tested rigorously.

Statistical hypothesis testing entails the comparison of observation
probabilities conditioned on different hypotheses
\cite{jaynes,vantrees}. To test two hypotheses labeled $\mathcal H_0$
and $\mathcal H_1$ using an observation record $Y$, the observer
splits the observation space into two parts $Z_0$ and $Z_1$; when $Y$
falls in $Z_0$, the observer chooses $\mathcal H_0$, and when $Y$
falls in $Z_1$, the observer chooses $\mathcal H_1$. The error
probabilities are then $P_{01} \equiv \int_{Z_0} dY P(Y|\mathcal H_1)$
and $P_{10} \equiv \int_{Z_1} dY P(Y|\mathcal H_0)$.  All binary
hypothesis testing protocols involve the computation of the likelihood
ratio, defined as
\begin{align}
\Lambda &\equiv \frac{P(Y|\mathcal H_1)}{P(Y|\mathcal H_0)}.
\label{like}
\end{align}
The ratio is then compared against a threshold $\gamma$ that depends on
the protocol; one decides on $\mathcal H_1$ if $\Lambda \ge \gamma$ and
$\mathcal H_0$ if $\Lambda <\gamma$. For example, the Neyman-Pearson
criterion minimizes $P_{01}$ under a constraint on $P_{10}$, while the
Bayes criterion minimizes $aP_{01}+bP_{10}$, $a$ and $b$ being
arbitrary positive numbers. For multiple independent trials, the final
likelihood ratio is simply the product of the ratios.

In most cases, the error probabilities are difficult to calculate
analytically and only bounds, such as the Chernoff upper bound
\cite{vantrees}, may be available, but the likelihood ratio can be
used to update the posterior hypothesis probabilities from prior
probabilities $P_0$ and $P_1$ via $P(\mathcal H_1|Y) = P_1\Lambda/(
P_1\Lambda + P_0)$ and $P(\mathcal H_0|Y) = P_0/( P_1\Lambda+P_0)$ and
therefore quantifies the strength of evidence for $\mathcal H_1$
against $\mathcal H_0$ given $Y$ \cite{jaynes}. Generalization to
multiple hypotheses beyond two is also possible by computing multiple
likelihood ratios or the posterior probabilities $P(\mathcal H_j|Y)$
\cite{vantrees}.

Consider now two hypotheses about a system under a sequence of
measurements, with results $ Y\equiv(\delta y_1,\dots,\delta y_M)$.
For generality, I use quantum theory to derive $P(Y|\mathcal H_j)$ for
both hypotheses, but note that a classical model can always be
expressed mathematically as a special case of a quantum model. The
observation probability distribution is \cite{wiseman}
\begin{align}
P(Y|\mathcal H_j) &= \trace \big[\mathcal J_j(\delta y_M,t_M)\mathcal K_j(t_M)
\dots
\nonumber\\&\quad
\mathcal J_j(\delta y_1,t_1)\mathcal K_j(t_1)\rho_j(t_0)\big],
\end{align}
where $\rho_j(t_0)$ is the initial density operator at time $t_0$,
$\mathcal K_j(t_m)$ is the completely positive map that models the
system dynamics from time $t_{m-1}$ to $t_m$, $\mathcal J_j(\delta
y_m,t_m)$ is the completely positive map that models the measurement
at time $t_m$, and the subscripts $j$ for $\rho_j(t_0)$, $\mathcal
K_j$, and $\mathcal J_j$ denote the assumption of $\mathcal H_j$ for
these quantities.

To proceed, let $t_m = t_0+m\delta t$ and assume the following Kraus
form of $\mathcal J_j$ for Gaussian measurements \cite{wiseman,tsang}:
\begin{align}
\mathcal J_j(\delta y,t)\rho &= \intall d(\delta z)
\frac{1}{\sqrt{2\pi S_j\delta t}}
\exp\Bk{-\frac{(\delta y-\delta z)^2}{2S_j\delta t}}
\nonumber\\&\quad\times
M_j(\delta z,t)\rho M_j^\dagger(\delta z,t),
\nonumber\\
M_j(\delta z,t) &\equiv
\frac{1}{(2\pi Q_j\delta t)^{1/4}}
\exp\bk{-\frac{\delta z^2}{4Q_j\delta t}}
\nonumber\\&\quad\times
\Bk{1+\frac{\delta z}{2Q_j}c_j-\frac{\delta t}{8Q_j}c_j^\dagger c_j
+o(\delta t)},
\end{align}
where $Q_j$ is the noise variance of the inherent quantum-limited
measurement, $S_j$ is the excess noise variance, $c_j$ is a quantum
operator depending on the measurement, and $o(\delta t)$ denotes terms
asymptotically smaller than $\delta t$.  
The map becomes
\begin{align}
\mathcal J_j(\delta y,t)\rho
&= \tilde P(\delta y)
\bigg[\rho + \frac{\delta y}{2R}\bk{c_j\rho+\rho c_j^\dagger}
\nonumber\\&\quad
+\frac{\delta t}{8Q_j}\bk{2c_j\rho c_j^\dagger - c_j^\dagger c_j\rho
-\rho c_j^\dagger c_j}+o(\delta t)\Big],
\nonumber\\
\tilde P(\delta y) &\equiv 
\frac{1}{\sqrt{2\pi R\delta t}}
\exp\bk{-\frac{\delta y^2}{2R\delta t}},
\quad
R \equiv Q_j + S_j.
\end{align}
I assume that the total noise variance $R$ is independent of the
hypothesis to focus on tests of hidden models rather than the
observation noise levels. $\tilde P(\delta y)$ then factors out of
both the numerator and denominator of the likelihood ratio and cancels
itself.

Taking the continuous time limit using It\={o} calculus with $\delta
y^2 =R\delta t + o(\delta t)$, the likelihood ratio becomes
$\Lambda = \trace f_1/\trace f_0$,
with $f_j$ obeying the following stochastic differential equation:
\begin{align}
df_j &= dt\mathcal L_j f_j + \frac{dy}{2R}\bk{c_jf_j + f_j c_j^\dagger}
\nonumber\\&\quad
+\frac{dt}{8Q_j}\bk{2c_jf_j c_j^\dagger - c_j^\dagger c_jf_j-f_j c_j^\dagger c_j}
\label{dmz}
\end{align}
and $\mathcal L_j$ being the Lindblad generator originating from
$\mathcal K_j$.  Equation~(\ref{dmz}) has the exact same mathematical
form as the linear Belavkin equation for an unnormalized filtering
density operator \cite{belavkin}, but beware that $f_j$ represents the
state of the system only if $\mathcal H_j$ is true; I call $f_j$ an
assumptive state.

To put $\Lambda$ in a form more amenable to numerics, consider the
stochastic differential equation for $\trace f_j$:
\begin{align}
d \trace f_j &= \trace df_j
= \frac{dy}{2R}\trace\bk{c_jf_j+f_jc_j^\dagger}
= \frac{dy}{R}\mu_j \trace f_j,
\label{dtrf}
\end{align}
where
\begin{align}
\mu_j &\equiv \frac{1}{\trace f_j}
\trace\bigg(\frac{c_j+c_j^\dagger}{2} f_j\bigg)
\label{estimate}
\end{align}
is an assumptive estimate; it is the posterior mean of the observable
$(c_j+c_j^\dagger)/2$ only if $\mathcal H_j$ is true. The form of
Eq.~(\ref{dtrf}) suggests that it can be solved by taking the
logarithm of $\trace f_j$, i.e., $d \ln \trace f_j = (d\trace
f_j)/\trace f_j -\bk{d\trace f_j}^2/2(\trace f_j)^2 = dy\mu_j/R -dt
\mu_j^2/2R$, resulting in
\begin{align}
\ln \trace f_j(T) &= 
\int_{t_0}^T\frac{dy}{R} \mu_j -\int_{t_0}^T\frac{dt}{2R} \mu_j^2,
\label{ito_integral}
\end{align}
with the $dy$ integral being an It\={o} integral.  $\Lambda$ becomes
\begin{align}
\Lambda(T) &= 
\exp\Bk{\int_{t_0}^T\frac{dy}{R}\bk{\mu_1-\mu_0} -\int_{t_0}^T\frac{dt}{2R} 
\bk{\mu_1^2-\mu_0^2}}.
\label{formula}
\end{align}
This compact formula for the likelihood ratio is the quantum
generalization of a similar result by Duncan and Kailath in classical
detection theory \cite{kailath}.  Generalization to the case of vectoral
observations with noise covariance matrix $R$ is trivial; the
result is simply Eq.~(\ref{formula}) with $dy\mu_j/R$ replaced by
$dy^\top R^{-1}\mu_j$ and $\mu_j^2/R$ by $\mu_j^\top R^{-1}\mu_j$.

For continuous measurements with Poissonian noise, a formula for
$\Lambda$ can be derived similarly \cite{sup}:
\begin{align}
\Lambda(T) &= \exp\Bk{\int_{t_0}^Tdy \ln \frac{\mu_1}{\mu_0}
-\int_{t_0}^T dt (\mu_1-\mu_0)},
\label{formula2}
\\
\mu_j &\equiv \frac{1}{\trace f_j}\trace(\eta_j c_j^\dagger c_j f_j),
\label{estimate2}
\\
df_j &=
dt\mathcal L_j f_j + (dy-\alpha dt)
\bk{\frac{\eta_j}{\alpha} c_jf_jc_j^\dagger -
f_j}
\nonumber\\&\quad
+ \frac{dt}{2} \bk{2c_jf_j c_j^\dagger-c_j^\dagger c_j f_j-f_jc_j^\dagger
c_j},
\label{filter2}
\end{align}
where $0<\eta_j\le 1$ is the quantum efficiency, $\alpha$ can be any
positive number, and Eqs.~(\ref{estimate2}) and (\ref{filter2}) form a
quantum filter for Poissonian observations
\cite{belavkin,tsang}. Equation~(\ref{formula2}) generalizes a similar
classical result by Snyder \cite{snyder}.

\begin{figure}[htbp]
\centerline{\includegraphics[width=0.4\textwidth]{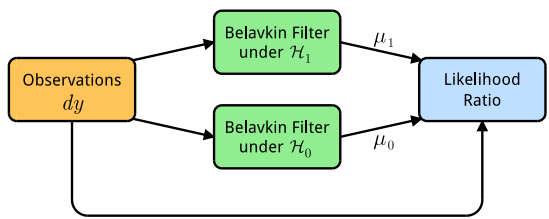}}
\caption{(Color online).  Structure of the likelihood-ratio formulae
  given by Eqs.~(\ref{formula}) and (\ref{formula2}).}
\label{estimator-correlator}
\end{figure}

Equations~(\ref{formula}) and (\ref{formula2}) show that continuous
hypothesis testing can be done simply by comparing how the observation
process is correlated with the observable estimated by each
hypothesis, as schematically depicted in
Fig.~\ref{estimator-correlator}.


Since Eqs.~(\ref{dmz}) and (\ref{estimate}) or Eqs.~(\ref{estimate2})
and (\ref{filter2}) have the same form as Belavkin filters, one can
leverage established quantum filtering techniques to update the
estimates and the likelihood ratio continuously with incoming
observations. If $f_j$ has a Wigner function that remains Gaussian in
time, the problem has an equivalent classical linear Gaussian model
\cite{wiseman,belavkin,tsang} conditioned on each hypothesis, and
$\mu_j$ can be computed efficiently using the Kalman-Bucy filter,
which gives the mean vector and covariance matrix of the Wigner
function. The classical model also enables one to use existing
formulae of Chernoff bounds for classical waveform detection
\cite{vantrees3} to bound the error probabilities.  It remains a
technical challenge to compute the quantum filter for problems without
a Gaussian phase-space representation beyond few-level systems, but
the quantum trajectory method should help cut the required
computational resources by employing an ensemble of wavefunctions
instead of a density matrix \cite{trajectory,wiseman}. Error bounds
for such nonclassical problems also remain an important open problem.

As an illustration of the theory, consider the detection of a weak
classical stochastic signal, such as a gravitational wave or a
magnetic field, using a quantum sensor \cite{braginsky}, with
$\mathcal H_1$ hypothesizing the presence of the signal and $\mathcal
H_0$ its absence. Let $x$ be a vector of the state variables for the
classical signal. One way to account for the dynamics of $x$ is to use
the hybrid density operator formalism, which includes $x$ as auxiliary
degrees of freedom in the system \cite{chase,tsang,warszawski}. The
initial assumptive state $f_1(t_0)$ becomes $\rho(t_0) P(x,t_0)$, with
$\rho(t_0)$ being the initial density operator for the quantum sensor
and $P(x,t_0)$ the initial probability density of
$x$. Equation~(\ref{dmz}) for $f_1$ becomes
\begin{align}
df_1 &= dt\mathcal L_1(x)f_1+
\frac{dy}{2R}\bk{cf_1 + f_1 c^\dagger}
\nonumber\\&\quad
+\frac{dt}{8Q}\bk{2cf_1 c^\dagger - c^\dagger cf_1-f_1 c^\dagger c},
\label{detection}
\end{align}
with $\mu_1 = \int dx \trace [(c+c^\dagger)f_1/2]/\int dx \trace f_1$.
$\mathcal L_1(x)$ should include the Lindblad generator for the
quantum sensor, the coupling of $x$ to the quantum sensor via an
interaction Hamiltonian, and also the forward Kolmogorov generator
that models the classical dynamics of $x$ \cite{tsang}. $c$ is an
operator that depends on the actual measurement of the quantum sensor;
for cavity optomechanical force detection for example, $c$ is the
cavity optical annihilation operator or can be approximated as the
mechanical position operator if the intracavity optical dynamics can
be adiabatically eliminated \cite{adiabatic}. 

For the null hypothesis $\mathcal H_0$, the classical degrees of
freedom need not be included. $f_0(t_0)$ is then $\rho(t_0)$,
Eq.~(\ref{dmz}) becomes
\begin{align}
df_0 &= dt\mathcal L_0f_0+
\frac{dy}{2R}\bk{cf_0 + f_0 c^\dagger}
\nonumber\\&\quad
+\frac{dt}{8Q}\bk{2cf_0 c^\dagger - c^\dagger cf_0-f_0 c^\dagger c},
\label{null}
\end{align}
and $\mathcal L_0$ includes only the Lindblad generator for the
quantum sensor. In most current cases of interest in quantum sensing,
the Wigner functions for $f_0$ and $f_1$ remain approximately Gaussian
\cite{braginsky,tsang}. Kalman-Bucy filters can then be used to solve
Eqs.~(\ref{detection}) and (\ref{null}) for the assumptive estimates,
to be correlated with the observation process according to
Eq.~(\ref{formula}) to produce $\Lambda$, and existing formulae of
Chernoff bounds for classial waveform detection \cite{vantrees3} can
be used to bound the error probabilities. Ref.~\cite{sup} contains a
simple example of such calculations.

Quantum smoothing can further improve the estimation of $x$
\cite{tsang} in the event of a likely detection.  Although smoothing
is not needed here for the exact computation of $\Lambda$, it may be
useful for improving the approximation of $\Lambda$ for non-Gaussian
problems when the exact estimates are too expensive to compute
\cite{crassidis}.

As a second example, consider the test of energy quantization in a
harmonic oscillator. To ensure the rigor of the test, imagine a
classical physicist who wishes to challenge the quantum harmonic
oscillator model by proposing a competing model based on classical
mechanics. To devise a good classical model, he first examines
quadrature measurements of a harmonic oscillator in a thermal
bath. With the harmonic time dependence on the oscillator frequency
removed in an interaction picture, the assumptive state $f_1$ for the
quantum hypothesis $\mathcal H_1$ obeys
\begin{align}
df_1 &= \frac{\gamma dt}{2}\big[\bk{N+1}
\bk{2af_1 a^\dagger -a^\dagger a f_1-f_1a^\dagger a}
\nonumber\\&\quad
+N\bk{2a^\dagger f_1 a -a a^\dagger f_1-f_1aa^\dagger}\big]
+\frac{dy}{2R}\bk{cf_1+f_1c}
\nonumber\\&\quad
+\frac{dt}{8Q}\bk{2cf_1c-c^2f_1-f_1c^2},
\label{ho}
\\
c &= q\cos\theta+p\sin\theta,
\end{align}
where $a \equiv (q+ip)/\sqrt{2}$ is the annihilation operator, $q$ and
$p$ are quadrature operators, $c$ is a quadrature operator with
$\theta$ held fixed for each trial to eliminate any complicating
measurement backaction effect, $\gamma$ is the decay rate of the
oscillator, and $N$ is a temperature-dependent parameter. This
backaction-evading measurement scheme can be implemented approximately
by double-sideband optical pumping in cavity optomechanics
\cite{braginsky,thorne}.

An equivalent classical model for the quadrature measurements is
\begin{align}
dx_1 &= -\frac{\gamma}{2} x_1 dt + \sqrt{\gamma} dW_1,
&dx_2 &= -\frac{\gamma}{2} x_2 dt + \sqrt{\gamma} dW_2,
\nonumber\\
dy &= h(x_1,x_2) dt + dV,
\label{x}
\\
h &= x_1\cos\theta+x_2\sin\theta.
\end{align}
where $x_1$ and $x_2$ are classical Ornstein-Uhlenbeck processes and
$dW_1$, $dW_2$, and $dV$ are uncorrelated classical Wiener noises with
$dW_1^2 = dW_2^2 =(N+1/2)dt$ and $dV^2 = Rdt$. One
can make $f_0$ diagonal and embed it with a classical distribution
$g_0(x_1,x_2)$ to model classical statistics; the equation for the
classical assumptive state $g_0$ is
\begin{align}
dg_0 &= \frac{\gamma dt}{2}
\bigg[\parti{}{x_1} \bk{x_1 g_0}+\parti{}{x_2} \bk{x_2 g_0}
\nonumber\\&\quad
+\bk{N+\frac{1}{2}}
\bk{\partit{g_0}{x_1}+\partit{g_0}{x_2}}\bigg]
+\frac{dy}{R}h g_0.
\label{g0}
\end{align}
which is a classical Duncan-Mortensen-Zakai (DMZ) equation \cite{dmz}.
The assumptive estimate $\mu_0 = \int dx_1 dx_2 h g_0/\int dx_1dx_2
g_0$ should be identical to the quantum one, as can be seen by
transforming $f_1$ to a Wigner function and neglecting the measurement
backaction that does not affect the observations. $\Lambda$ given the
quadrature observations then stays at 1, confirming that the two
models are indistinguishable.

In a different experiment on the same oscillator, the energy of the
oscillator is measured instead. Let
\begin{align}
c = \frac{q^2+p^2}{2} = \frac{a^\dagger a + a a^\dagger}{2},
\label{energy_c}
\end{align}
which can be implemented approximately by dispersive optomechanical
coupling in cavity optomechanics \cite{braginsky,thompson,santamore}.
$f_1$ still obeys Eq.~(\ref{ho}), but with $c$ now given by
Eq.~(\ref{energy_c}) and different $R$ and $Q$.  The measurements are
again backaction-evading, as the backaction noise on the oscillator
phase does not affect the energy observations.

Given the prior success of the classical model, the classical
physicist decides to retain Eqs.~(\ref{x}) and modifies only the
observation as a function of $x_1$ and $x_2$:
\begin{align}
h &= \frac{x_1^2+x_2^2}{2}.
\end{align}
The DMZ equation given by Eq.~(\ref{g0}), assuming continuous energy,
should now produce an assumptive energy estimate different from the
quantum one; it is this difference that should make the likelihood
ratio increase in favor of the quantum hypothesis with more
observations, if quantum mechanics is correct.  Previous data analysis
techniques that consider only the quantum estimate \cite{santamore}
fail to take into account the probability that the observations can
also be explained by a continuous-energy model and are therefore
insufficient to demonstrate energy quantization conclusively.  The
non-Gaussian nature of the problem means that bounds on the error
probabilities may be difficult to compute analytically and one may
have to resort to numerics, but one can also use $\Lambda$ as a
Bayesian statistic to quantify the strength of the evidence for one
hypothesis against another \cite{jaynes}.


Discussions with J.~Combes, C.~Caves, and A.~Chia are gratefully
acknowledged. This material is based on work supported by the
Singapore National Research Foundation under NRF Grant
No. NRF-NRFF2011-07.

\appendix
\begin{widetext}
\section{Supplementary Material}
This document contains a derivation of the likelihood-ratio formula
for continuous quantum measurements with Poissonian noise in
Sec.~\ref{poisson} and an example of quantum optomechanical
stochastic force detection in Sec.~\ref{force}.

\subsection{\label{poisson}Likelihood-ratio formula for continuous Poissonian
measurements}
The completely positive map for a weak Poissonian measurement
is given by
\begin{align}
\mathcal J_j(\delta y)\rho
&= 
 \sum_{\delta z = 0,1} P(\delta y|\delta z)
\BK{\delta z c_j\rho c_j^\dagger \delta t
+(1-\delta z)\Bk{\rho-
\frac{\delta t}{2}(c_j^\dagger c_j\rho+\rho c_j^\dagger c_j)}},
\end{align}
where $\delta y,\delta z \in \{0,1\}$ and
\begin{align}
P(\delta y|\delta z) = (1-\delta y)(1-\eta_j\delta z)+\eta_j\delta y \delta z
\end{align}
models the effect of imperfect quantum efficiency $0<\eta_j\le 1$.
Rearranging terms \cite{tsang}, 
\begin{align}
\mathcal J_j(\delta y)\rho
&= \tilde P(\delta y)
\Bk{\rho +\frac{\delta t}{2}\bk{2c_j\rho c_j^\dagger
-c_j^\dagger c_j\rho -\rho c_j^\dagger c_j}
+(\delta y-\alpha\delta t)\bk{\frac{\eta_j}{\alpha}c_j\rho c_j^\dagger
-\rho}},
\\
\tilde P(\delta y) &\equiv (1-\delta y)(1-\alpha\delta t)+
\delta y\alpha\delta t,
\end{align}
where $\tilde P(\delta y)$ is a reference probability distribution and
$\alpha$ is an arbitrary positive number.
This gives
\begin{align}
\Lambda &= \frac{\trace f_1}{\trace f_0},
\\
df_j &= dt\mathcal L_j f_j + \frac{dt}{2}\bk{2c_jf_jc_j^\dagger
-c_j^\dagger c_jf_j-f_j c_j^\dagger c_j}
+(dy-\alpha dt)\bk{\frac{\eta_j}{\alpha}c_jf_jc_j^\dagger - f_j}.
\label{df}
\end{align}
Equation~(\ref{df}) coincides with the quantum filtering equation
for an unnormalized posterior density operator $f_j$ given
Poissonian observations $dy$ \cite{tsang}. Next, consider
\begin{align}
d\trace f_j &= \trace df_j = (dy-\alpha dt)
\bk{\frac{\mu_j}{\alpha}-1}\trace f_j,
\\
\mu_j &\equiv \frac{1}{\trace f_j}\trace\bk{\eta_j c_j^\dagger c_j f_j},
\end{align}
where $\mu_j$ is the filtering estimate of the observable $\eta_j
c_j^\dagger c_j$ assuming that the hypothesis $\mathcal H_j$ is
true. Expanding $d\ln \trace f_j$ in Taylor series,
\begin{align}
d\ln \trace f_j &= 
\sum_{n = 1}^\infty \frac{(-1)^{n+1}}{n(\trace f_j)^n} (d\trace f_j)^n
\\
&= \sum_{n = 1}^\infty \frac{(-1)^{n+1}}{n}(dy-\alpha dt)^n
\bk{\frac{\mu_j}{\alpha}  - 1}^n.
\end{align}
With $(dy-\alpha dt)^n = dy^n + o(dt)$ for $n \ge 2$ and $dy^n = dy$
for Poissonian observations,
\begin{align}
d\ln \trace f_j &= 
dy\ln \frac{\mu_j}{\alpha}- dt\bk{\mu_j  - \alpha},
\\
\ln \trace f_j(T) &= 
\int_{t_0}^T dy\ln \frac{\mu_j}{\alpha}- \int_{t_0}^T
dt\bk{\mu_j  - \alpha},
\\
\Lambda(T) &= 
\exp\Bk{\int_{t_0}^T dy\ln \frac{\mu_1}{\mu_0}- \int_{t_0}^T
dt\bk{\mu_1  - \mu_0}}.
\end{align}

\subsection{\label{force} Quantum optomechanical detection
of a Gaussian stochastic force}
Let $F = C x$ be a classical force acting on a moving mirror with
position operator $q$ and momentum $p$, and $x$ be a vectoral
classical Gaussian stochastic process $x$ described by the Ito
equation
\begin{align}
dx = Axdt + dW,\quad
dW dW^\top  = Bdt.
\end{align} 
The mirror is
assumed to be a harmonic oscillator with mass $m$ and frequency
$\omega$ and part of an optical cavity pumped by a near-resonant
continuous-wave laser. The phase quadrature of the cavity output is
measured continuously by homodyne detection, with an observation
process given by $dy$.  For simplicity, I assume that the optical
intracavity dynamics can be adiabatically eliminated, the phase
modulation by the mirror motion is much smaller than $\pi/2$ radians,
such that the homodyne detection is effectively measuring the mirror
position, and there is no excess decoherence. Under hypothesis
$\mathcal H_1$, the force is present and the quantum filtering
equation for the unnormalized hybrid density operator $f_1(x,t)$ is
then given by \cite{tsang}
\begin{align}
df_1 &= dt\mathcal L_1(x) f_1 + \frac{dy}{2R}\bk{q f_1+ f_1 q}
+\frac{dt}{8R}\bk{2qf_1q-q^2f_1-f_1q^2},
\\
\mathcal L_1(x)f_1 &\equiv -\frac{i}{\hbar}[H_1(x),f_1] +\mathcal L_c(x) f_1,
\\
H_1(x) &\equiv \frac{p^2}{2m} +\frac{m\omega^2}{2}q^2-q Cx,
\\
\mathcal L_c(x) f_1 &\equiv -\sum_\mu \parti{}{x_\mu}\Bk{(A x)_\mu f_1}
+\frac{1}{2}\sum_{\mu,\nu} \parti{^2}{x_\mu\partial x_\nu}
\bk{B_{\mu\nu}f_1},
\end{align}
where $R$ is the measurement noise variance that depends on the laser
intensity and the cavity properties and $\mathcal L_c(x)$ is the forward
Kolmogorov generator for the classical process $x$. Under
the null hypothesis $\mathcal H_0$, the force is absent and the
filtering equation for the oscillator density operator $f_0$ is
\begin{align}
df_0 &= dt\mathcal L_0 f_0 + \frac{dy}{2R}\bk{q f_0+ f_0 q}
+\frac{dt}{8R}\bk{2qf_0q-q^2f_0-f_0q^2},
\\
\mathcal L_0 f_0 &\equiv -\frac{i}{\hbar}[H_0,f_0] ,
\\
H_0 &\equiv \frac{p^2}{2m} +\frac{m\omega^2}{2}q^2.
\end{align}
These filtering equations can be transformed to equations
for the Wigner functions of $f_j$:
\begin{align}
g_1(q,p,x,t) &\equiv \frac{1}{2\pi \hbar}
\intall du \bra{q-u/2}f_1(x,t)\ket{q+u/2}\exp(ip u/\hbar),
\\
g_0(q,p,t) &\equiv \frac{1}{2\pi \hbar}
\intall du \bra{q-u/2}f_0(t)\ket{q+u/2}\exp(ip u/\hbar),
\\
dg_j &= dt \mathcal L_j' g_j + \frac{dy}{R} qg_j,
\\
\mathcal L_1' g_1 &\equiv \mathcal L_c(x) g_1 + dt
\Bk{-\frac{p}{m}\parti{g_1}{q}+(m\omega_m^2 q-Cx)\parti{g_1}{p}+
\frac{\hbar^2}{8R}\partit{g_1}{p}},
\\
\mathcal L_0' g_0 &\equiv 
-\frac{p}{m}\parti{g_0}{q}+m\omega^2 q\parti{g_0}{p}
+\frac{\hbar^2}{8R}\partit{g_0}{p},
\end{align}
where $q$ and $p$ are now phase-space variables, $p$ is seen to suffer
from measurement-back-action-induced diffusion, and $\mathcal L_1'$
has the form of a forward Kolmogorov generator for a new Gaussian
process $z = (q,p,x^\top )^\top $:
\begin{align}
\mathcal L_1' g_1(z,t) &= 
-\sum_\mu \parti{}{z_\mu}\Bk{(J_1 z)_\mu g_1}
+\frac{1}{2}\sum_{\mu,\nu} \parti{^2}{z_\mu\partial z_\nu}
\bk{S_{1\mu\nu}g_1},
\\
J_1 &\equiv \bk{\begin{array}{cc|c}0 & 1/m & \bs 0 \\
-m\omega^2 & 0 & C\\
\hline
\bs 0 & \bs 0 & A \end{array}},
\quad
S_1 \equiv 
\bk{\begin{array}{cc|c}0 & 0 & \bs 0 \\
0 & \hbar^2/4R & \bs 0\\
\hline
\bs 0 & \bs 0 & B \end{array}},
\end{align}
with $\bs 0$ denoting zero matrices.
Similarly, under $\mathcal H_0$, we have $z= (q,p)^\top $ and
\begin{align}
\mathcal L_0' g_0(z,t) &= 
-\sum_\mu \parti{}{z_\mu}\Bk{(J_0 z)_\mu g_0}
+\frac{1}{2}\sum_{\mu,\nu} \parti{^2}{z_\mu\partial z_\nu}
\bk{S_{0\mu\nu}g_0},
\\
J_0 &\equiv \bk{\begin{array}{ccc}0 & 1/m  \\
-m\omega^2 & 0 \end{array}},
\quad
S_0 \equiv 
\bk{\begin{array}{ccc}0 & 0  \\
0 & \hbar^2/4R \end{array}}.
\end{align}
The Gaussian statistics mean that we can use Kalman-Bucy filters to
compute the filtering estimates of the mirror position $\mu_j$ given
$dy$ \cite{tsang}:
\begin{align}
dz_j' &= J_jz_j'dt +\Gamma_j(dy-K_jz_j'dt),
\quad
K_j \equiv (1,0,\dots,0),
\\
\Gamma_j &\equiv \Sigma_jK_j^\top  R^{-1},
\\
\diff{\Sigma_j}{t} &= J_j\Sigma_j+\Sigma_j J_j^\top -\Sigma_j K_j^\top R^{-1}K_j\Sigma_j^\top 
+S_j,
\label{sigma}
\\
\mu_j &= K_j z_j',
\end{align}
and the likelihood ratio becomes
\begin{align}
\Lambda(T) &= \exp\Bk{\int_{t_0}^T \frac{dy}{R}(\mu_1-\mu_0)-\int_{t_0}^T
\frac{dt }{2R}(\mu_1^2-\mu_0^2)}.
\end{align}
Given the Gaussian structure of the problem under each hypothesis, we
can use known results about the Chernoff upper bounds for classical
waveform estimation to bound the error probabilities \cite{vantrees3}:
\begin{align}
P_{10} &\le \exp\Bk{\mu(s)-s\gamma},
\\
P_{01} &\le \exp\Bk{\mu(s)+(1-s)\gamma},
\\
\mu(s) &= \frac{1}{2R}\int_{t_0}^T dt
\Bk{(1-s)\Sigma_{1q}(t)+s\Sigma_{0q}(t)
-\tilde\Sigma_{q}(s,t)},
\end{align}
where $0\le s\le 1$, $\gamma$ is the threshold of the likelihood-ratio
test, $\Sigma_{jq}(t)$ is the $q$ variance component of $\Sigma_j$,
which obeys Eq.~(\ref{sigma}), and $\tilde\Sigma_q(s,t)$ is the
variance of $\sqrt{s}q_0 + \sqrt{1-s}q_1$ for a different filtering
problem, in which observations of $\sqrt{s}q_0 + \sqrt{1-s}q_1$ are
made with noise variance $R$ and $q_j$ has the statistics of $q$ under
$\mathcal H_j$, viz.,
\begin{align}
\diff{\tilde\Sigma}{t} &= \tilde J\tilde \Sigma+\tilde \Sigma \tilde J^\top -
\tilde\Sigma\tilde K^\top R^{-1}\tilde K\tilde\Sigma^\top +\tilde S,
\\
\tilde J &\equiv \bk{\begin{array}{c|c}J_0 & \bs 0\\
\hline
\bs 0 & J_1\end{array}},
\quad \tilde S  \equiv 
\bk{\begin{array}{c|c}S_0 & \bs 0\\
\hline
\bs 0 & S_1\end{array}},
\quad
\tilde K \equiv \bk{
\begin{array}{cc|cc}\sqrt{s} &\bs 0 & \sqrt{1-s} &\bs 0\end{array}},
\quad
\tilde\Sigma_q \equiv \tilde K\tilde\Sigma\tilde K^\top .
\end{align}
The tightest upper bounds are obtained by minimizing the bounds with
respect to $s$. If $x$ and therefore $q$ are stationary,
$\Sigma_j$ and $\tilde\Sigma$ will converge to steady states
in the long-time limit, and the Chernoff bounds will decay
exponentially with time. 

\end{widetext}
\end{document}